\documentclass[trackchanges, twocolumn]{aastex701}

\shorttitle{Tidal Tails of NGC~5024}
\shortauthors{Ye et al.}
\received{xxx xx, 2026}
\revised{xxx xx, 2026}
\accepted{xxx xx, 2026}
\submitjournal{ApJL}

\begin{document}

\title{Long Tidal Tails of NGC~5024 Hidden in LMS-1 and NGC~5053 Tidal Streams}

\author[orcid=0000-0002-5805-8112]{Xianhao Ye}
\affiliation{Institute of Astronomy and Physics, Inner Mongolia University, Hohhot 010021, People's Republic of China}
\affiliation{National Astronomical Observatories, Chinese Academy of Sciences, Beijing 100101, People's Republic of China}
\affiliation{Instituto de Astrof\'{i}sica de Canarias, V\'{i}a L\'{a}ctea, 38205 La Laguna, Tenerife, Spain}
\affiliation{Universidad de La Laguna, Departamento de Astrof\'{i}sica, 38206 La Laguna, Tenerife, Spain}
\email[show]{xhye@imu.edu.cn}

\author[0000-0001-7609-1947]{Yong Yang}
\affiliation{Institute of Astronomy and Physics, Inner Mongolia University, Hohhot 010021, People's Republic of China}
\email{yong.yang@sydney.edu.au}

\author[orcid=0000-0003-2868-8276]{Jingkun Zhao}
\affiliation{National Astronomical Observatories, Chinese Academy of Sciences, Beijing 100101, People's Republic of China}
\affiliation{Institute of Astronomy and Physics, Inner Mongolia University, Hohhot 010021, People's Republic of China}
\email[show]{zjk@nao.cas.cn}

\author[orcid=0000-0003-3347-7596]{Hao Tian}
\affiliation{National Astronomical Observatories, Chinese Academy of Sciences, Beijing 100101, People's Republic of China}
\affiliation{Institute for Frontiers in Astronomy and Astrophysics, Beijing Normal University, Beijing 102206, People's Republic of China}
\email{tianhao@nao.cas.cn}

\author[orcid=0000-0002-8442-901X]{Yuqin Chen}
\affiliation{National Astronomical Observatories, Chinese Academy of Sciences, Beijing 100101, People's Republic of China}
\affiliation{Institute of Astronomy and Physics, Inner Mongolia University, Hohhot 010021, People's Republic of China}
\email{cyq@nao.cas.cn}

\author[orcid=0000-0002-8980-945X]{Gang Zhao}
\affiliation{National Astronomical Observatories, Chinese Academy of Sciences, Beijing 100101, People's Republic of China}
\affiliation{Institute of Astronomy and Physics, Inner Mongolia University, Hohhot 010021, People's Republic of China}
\email[show]{gzhao@nao.cas.cn}

\correspondingauthor{Xianhao Ye}
\correspondingauthor{Jingkun Zhao}
\correspondingauthor{Gang Zhao}


\begin{abstract}
     
  We report the discovery of long tidal tails associated with the globular cluster (GC) NGC~5024. A modified matched filter applied to Gaia DR3 data reveals a broad stellar stream spanning $\alpha \approx 230^{\circ}-175^{\circ}$. The stellar stream overlaps on the sky with the LMS-1 and with the simulated stream of the GC NGC~5053, and all three share similar proper motions and metallicity. Our member candidates may therefore be a mixture of stars from NGC~5024, NGC~5053, and LMS-1. Nevertheless, the trailing tail extends roughly $20^{\circ}$ beyond the known LMS-1 stream. Furthermore, the radial velocity (RV) as a function of $\alpha$ is used to distinguish the genuine stream candidates of NGC~5024 from the streams of NGC~5053 and LMS-1. Among the sources in common with DESI (Dark Energy Spectroscopic Instrument) DR1, we found two distinct sequences in RV-$\alpha$ plane, corresponding to the stellar streams of NGC~5024 and NGC~5053 (or LMS-1), respectively. This constitutes the first strong evidence for the existence of extensive tidal tails around NGC~5024.

\end{abstract}


\section{Introduction}\label{sec:intro}

The Galactic halo has experienced multiple merger events and is rich in substructures and stellar streams from these accretions \citep{1994Natur.370..194I,2003ApJ...585L.125B,2010MNRAS.408L..26P,2018MNRAS.478..611B,2018Natur.563...85H,2019MNRAS.486.3180K,2019NatAs...3..667I,2019MNRAS.488.1235M,2019A&A...631L...9K,2020MNRAS.493.3363H,2020ApJ...901...48N,2022MNRAS.509.5992S,2022ApJ...930L...9M}. Unlike debris from dwarf galaxies, stellar streams originating from the globular clusters (GCs) are usually dynamically cold and thin \citep[e.g.,][]{2010ApJ...712..260K,2020MNRAS.493.4978S,2021ApJ...914..123I}.

\begin{deluxetable*}{cccccccccc} 
\tablewidth{0pt}
\tablecaption{Adopted parameters for NGC~5024 and corresponding literature references. \label{t:parameters}}
\tablehead{
    R.A. (J2000) & Decl. (J2000) & $\mu_{\alpha}^{*}$ & $\mu_{\delta}$ & $D$ & RV & [Fe/H] & Age & Mass & $r_{\mathrm{hm}}$ \\
    deg & deg & mas $\cdot$ yr$^{-1}$ & mas $\cdot$ yr$^{-1}$ & kpc & km $\cdot$ s$^{-1}$ & dex & Gyr & $\mathrm{M_{\odot}}$ & pc
}
\startdata
198.230 & 18.168 & -0.131 & -1.332 & 18.50 & -63.37 & -2.0 & 12.7 & 380000 & 9.92 \\
\enddata
\tablecomments{The right ascension (R.A.), declination (Decl.), proper motions $\left( \mu_{\alpha}^{*}, \mu_{\delta} \right)$, distance $D$, and radial velocity (RV) are taken from \cite{2021MNRAS.505.5978V,2021MNRAS.505.5957B}. The metallicity $\mathrm{[Fe/H]}=-2.0$ is adopted from \cite{2020ApJ...900..146C}. We adopt an age of $12.7$ Gyr \citep{2021AJ....162...42W}, while the total mass and half-mass radius $r_{\mathrm{hm}}$ of this cluster are taken from \cite{2018MNRAS.478.1520B}.}
\end{deluxetable*}

In the past few decades, efforts have been made to search for member stars of NGC~5024 tidal tails. For example, \cite{2010A&A...522A..71J,2019MNRAS.483.1737K,2020ApJ...900..146C} report that an extratidal halo exists around NGC~5024 beyond its tidal radius, and this cluster might also be related to the Sagittarius dwarf galaxy \citep{2010A&A...522A..71J}. However, a later study by \cite{2020AJ....160...31N}, which, like \cite{2019MNRAS.483.1737K}, used RR Lyrae stars to probe NGC~5024, find that no RR Lyrae belonging to NGC~5024 are extra-tidal. Moreover, \cite{2020ApJ...900..146C} suggest that the GCs NGC~5024 and NGC~5053 are tidally interacted. These two clusters lies close together on the sky and have similar proper motions. In the related studies, \cite{2020ApJ...898L..37Y} find that the LMS-1 stream, which is linked to a dwarf galaxy, is very similar to NGC~5024 and NGC~5053 in orbital parameters and metallicity \citep{2021ApJ...920...51M}. In addition, \cite{2021ApJ...909L..26B} find in orbital phase-space that NGC~5024 is possibly the progenitor of two stellar streams, Sylgr and Ravi. More recently, \cite{2025MNRAS.540.2863W} also claim that a tidal structure exists around this cluster. By studying the stellar distribution within $10^{\circ}$ of NGC~5053, \cite{2026ApJS..283...60C} report that both the stream of NGC~5053 and that of NGC~5024 should exist.

In this study, we revisit the candidate stars tidally stripped from NGC~5024 using Gaia DR3 data, and we find strong evidence for the existence of a long stellar stream associated with NGC~5024. \S\ref{sec:data} describes how we construct our data sample from Gaia. We present the methods we used to identify the tidal tails in \S\ref{sec:methods}, discuss the results in \S\ref{sec:results}, and summarize the paper in \S\ref{sec:summary}.

\section{Data}\label{sec:data}

The tidal tails associated with NGC~5024 are identified by combing Gaia DR3 astrometric and photometric data with mock streams. We apply the following criteria to primarily select stars within a region encompassing the core of NGC~5024 and its potential tidal structures. In addition, we follow the calculations in \cite{2021A&A...649A...3R} to derive the corrected BP and RP flux excess factor $C^{*}$ for selecting stars with reliable photometry, and we apply a cut of $\texttt{RUWE} < 1.4$ to remove sources with unreliable astrometric solution.
\begin{itemize}
  \item Remove disk stars: $|b| > 40^{\circ}$,
  \item Remove foreground stars: $1/\varpi > 5$ kpc,
  \item $175^{\circ} < \mathrm{R.A.} < 230^{\circ}$, $-10^{\circ} < \mathrm{Decl.} < 30^{\circ}$,
  \item $|C^{*}| < 3 \sigma_{C^{*}}$
  \item $\texttt{RUWE} < 1.4$ .
\end{itemize}
Furthermore, we remove some high-density clusters from the sample, as they could interfere with the identification of the relatively low-density tidal tails. We roughly estimate the centers and radii of these clusters using \texttt{TOPCAT} and discard all stars falling within these regions. The adopted centers and radii are listed below, all in degrees.
\begin{itemize}
  \item \texttt{R.A.}, \texttt{Decl.}, \texttt{radius} = 199.113, 17.700, 0.06
  \item \texttt{R.A.}, \texttt{Decl.}, \texttt{radius} = 205.551, 28.374, 0.25
  \item \texttt{R.A.}, \texttt{Decl.}, \texttt{radius} = 211.366, 28.532, 0.08
  \item \texttt{R.A.}, \texttt{Decl.}, \texttt{radius} = 229.641, 2.080, 0.19
  \item \texttt{R.A.}, \texttt{Decl.}, \texttt{radius} = 217.404, -5.975, 0.03
  \item \texttt{R.A.}, \texttt{Decl.}, \texttt{radius} = 182.523, 18.542, 0.03
\end{itemize}
Finally, the sample is restricted to stars with $16.5 < G_{0} < 20$ mag both to minimize contamination by bright sources and to ensure reliable photometry. The extinction-corrected magnitudes are computed assuming $A_{G}/A_{V}=0.83627$, $A_{G_{\mathrm{BP}}}/A_{V}=1.08337$, and $A_{G_{\mathrm{RP}}}/A_{V}=0.63439$\footnote{These ratios can be obtained from the PARSEC website: \url{https://stev.oapd.inaf.it/cgi-bin/cmd_3.8}}. The extinction $A_{V}$ for each star is estimated from the two-dimensional dust map \citep{1998ApJ...500..525S,2011ApJ...737..103S}, accessed through the \texttt{dustmaps} \texttt{Python} package \citep{2018JOSS....3..695G}, adopting $R_{V}=3.1$. After applying all criteria, our final sample contains $500,950$ sources in the field of NGC~5024.

\section{Methods}\label{sec:methods}

We follow the modified matched filter method described in \cite{2019ApJ...884..174G,2022ApJ...935L..38Y,2023ApJ...953..130Y} to estimate the membership weight for each star. In brief, a star that is closer to the theoretical isochrone of NGC~5024 in a color-magnitude diagram (CMD), and whose proper motions are more consistent with the mock stream will be assigned a larger weight.

\begin{figure*}[htbp!]
  \centering
  \includegraphics[width = 0.49\textwidth]{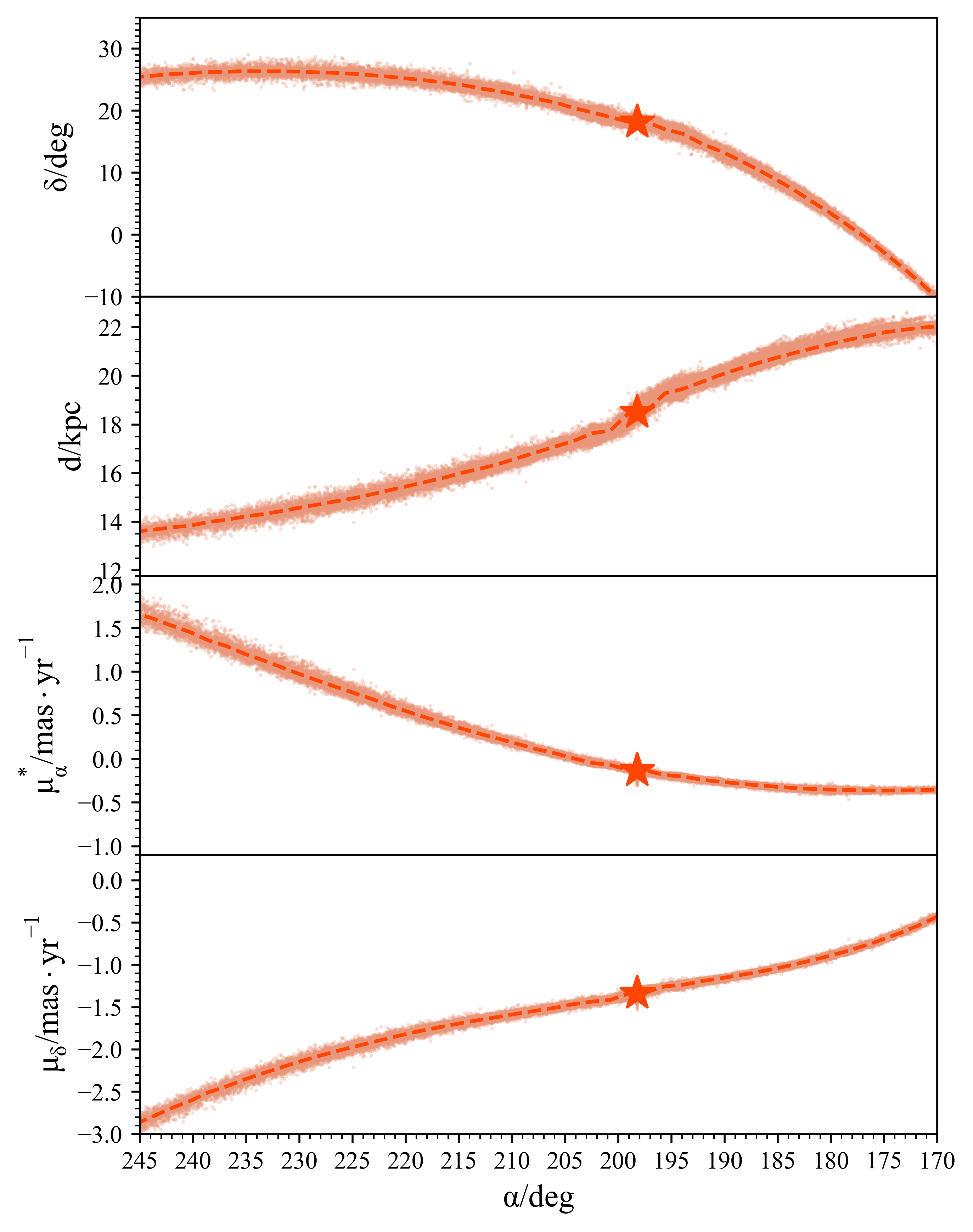}
  \includegraphics[width = 0.44\textwidth]{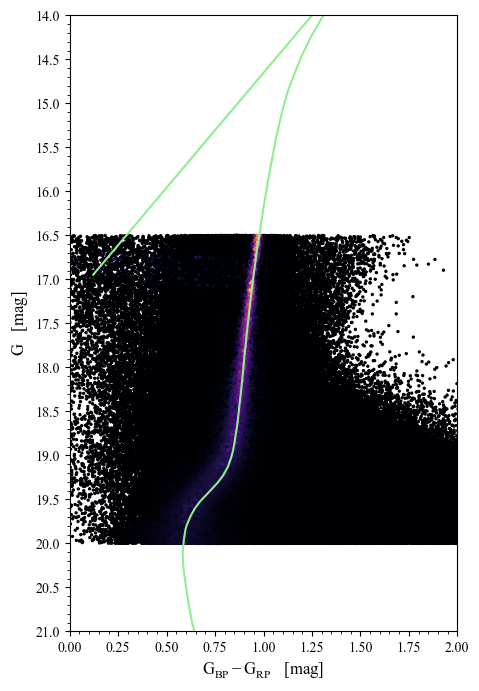}
  \caption{Simulated tidal streams of NGC~5024 and Gaia CMD. Left panel: NGC~5024 and its mock stream shown in different projections of phase-space. The dashed lines indicate the median values of $\delta$, $D$, $\mu_{\alpha}^{*}$, $\mu_{\delta}$ as a function of $\alpha$. Right panel: CMD using Gaia photometry. The color scale encodes the weight of each Gaia source according to the isochrone, where the PARSEC isochrone for NGC~5024 is overplotted as the green line.}
  \label{fig:cmd_mock}
\end{figure*}

The basic parameters adopted for NGC~5024, compiled from the literature, are listed in Table~\ref{t:parameters}. We extract a $12.7$ Gyr isochrone with $\mathrm{[Fe/H]}=-2.0$ from the PARSEC version 1.2S stellar tracks \citep{2012MNRAS.427..127B}. We use \texttt{Gala}\footnote{\url{https://gala.adrian.pw/en/latest/}} \citep{gala,2024zndo..10449846P} to generate a mock stream for NGC~5024. The simulation adopts the default static potential \texttt{MilkyWayPotential} in \texttt{Gala}, and includes the gravitational influence of the Large Magellanic Cloud (LMC), modeled as described in \cite{2022MNRAS.513..853Y}. All other details of the mock-stream generation are the same as in \cite{2022MNRAS.513..853Y,2022ApJ...935L..38Y}. In Figure~\ref{fig:cmd_mock}, we present the CMD and the mock stream in different phase-space projections. The stream is initiated $3$ Gyr ago by releasing particles from the leading and trailing arms \citep{2014MNRAS.445.3788G,2019MNRAS.487.2685E}. The isochrone is placed at a distance of $D=18.5$ kpc.

\begin{figure*}[htbp!]
  \centering
  \includegraphics[width = 1.00\textwidth]{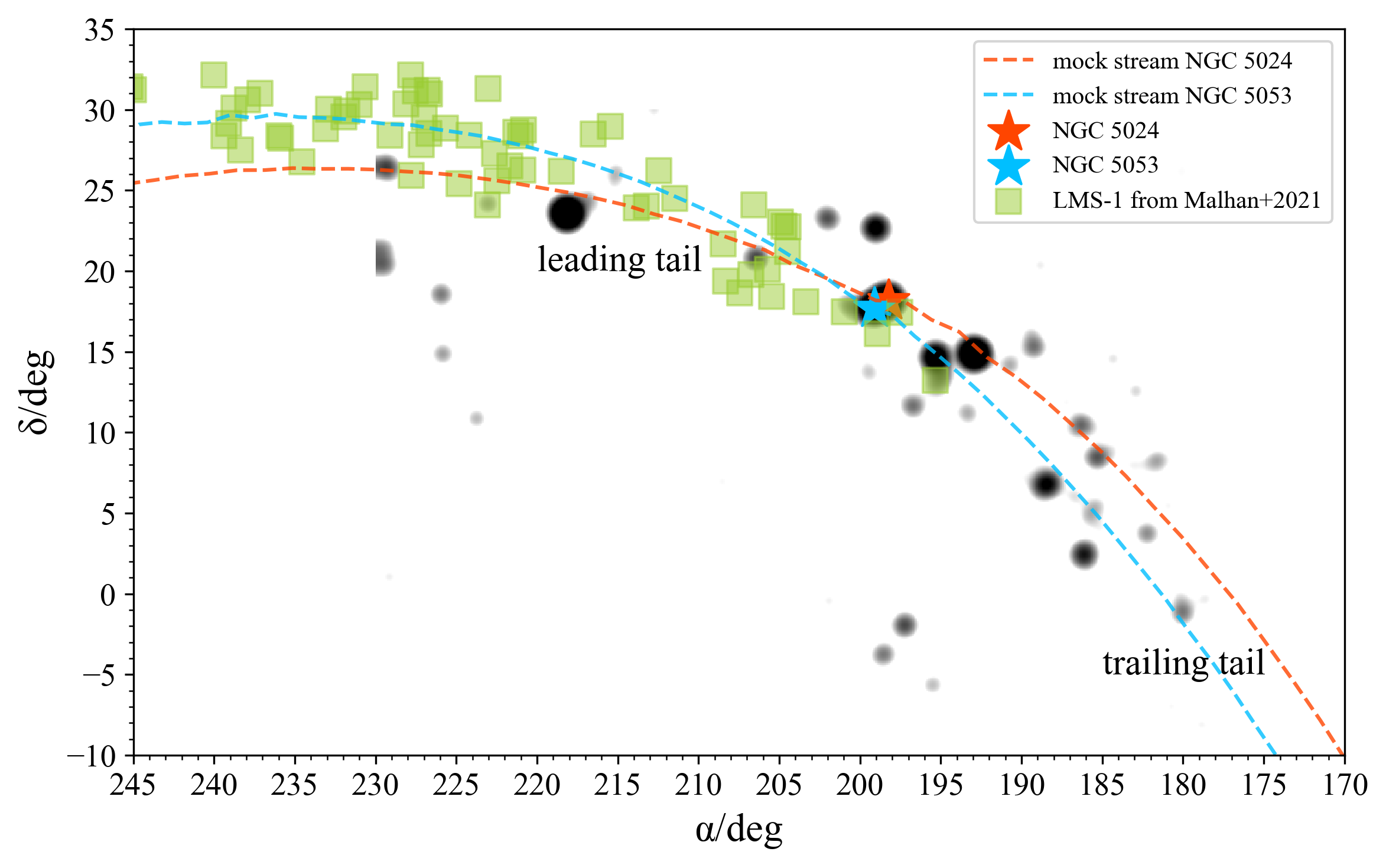}
  \caption{Weighted sky map of our sample, smoothed with a Gaussian filter ($\sigma=5$). The image is scaled between $0.7$ to $5$ (using \texttt{vmin} and \texttt{vmax} parameters of \texttt{matplotlib.imshow}) to enhance the visibility of the leading and trailing tails. The position of NGC~5024 and the median location of its simulated tidal stream are marked in red, while NGC~5053 and its stream median are shown in blue. The stellar stream LMS-1 is also plotted, adopted from \cite{2021ApJ...920...51M}.}
  \label{fig:weight_skymap}
\end{figure*}

\section{Results}\label{sec:results}

After applying our modified matched filter to the stellar sample, a long and relatively broad stellar stream extending from $\alpha \sim 230^{\circ}$ to $\alpha \sim 180^{\circ}$ is clearly visible, as shown in the weighted map of Figure~\ref{fig:weight_skymap}. The median paths of two model streams from NGC~5024 and NGC~5053 are indicated by dashed lines. The adopted parameters of NGC~5053 are taken from the same references as those listed in Table~\ref{t:parameters}. These two mock streams are very close in $\alpha$-$\delta$ diagram, particularly when their widths are taken into account. Moreover, they remain close in the phase-spaces projections $\alpha$, $\delta$, $\mu_{\alpha}^{*}$, $\mu_{\delta}$.

The LMS-1 stellar stream also follows a similar track. Its member stars \citep{2021ApJ...920...51M} are highlighted with green squares in Figure~\ref{fig:weight_skymap} for comparison. As shown in the figure, the predicted path of NGC~5053 leading tail overlaps with the LMS-1 stream members identified in \cite{2021ApJ...920...51M}. According to \cite{2021ApJ...920...51M}, the LMS-1 stream and the tidal stream of NGC~5053 are remarkably similar in all projection phase-spaces, including $\alpha$, $\delta$, $\mu_{\alpha}^{*}$, $\mu_{\delta}$, and even RV.

\subsection{Stream member candidates of NGC~5024}

Given the width of the stream, the feature we detected may consist of not only the tidal tails of NGC~5024, but also those of NGC~5053 and LMS-1. In the next subsection, we present evidence for the genuine stream members of NGC~5024. In this subsection, we first identify the member candidates along the tracks of the mock stellar streams for NGC~5024 and NGC~5053. A trailing tail, approximately $20^{\circ}$ long in the $\alpha$ direction, is revealed beyond the cores of these two GCs and the LMS-1 stream reported in \cite{2021ApJ...920...51M}. We select the $51$ candidates with the highest weights from Figure~\ref{fig:weight_skymap} and plot them in the phase-space projections shown in Figure~\ref{fig:members}. We note, however, that contamination from LMS-1 and NGC~5053 stream stars is likely present among these candidates. These stars are listed in Appendix~\ref{sec:appendix}. The extracted stream candidates closely follow the track of the mock stream of NGC~5024 in all projection phase-spaces and agree with the PARSEC isochrone in the CMD. For comparison, the CMD in Figure~\ref{fig:members} includes the LMS-1 and NGC~5053 stream members taken from \cite{2021ApJ...920...51M,2026ApJS..283...60C}. Our proposed stream candidates trace a CMD locus consistent with that of the literature stars.

\subsection{RV of the member candidates}

\begin{figure*}[htbp!]
  \centering
  \includegraphics[width = 1.00\textwidth]{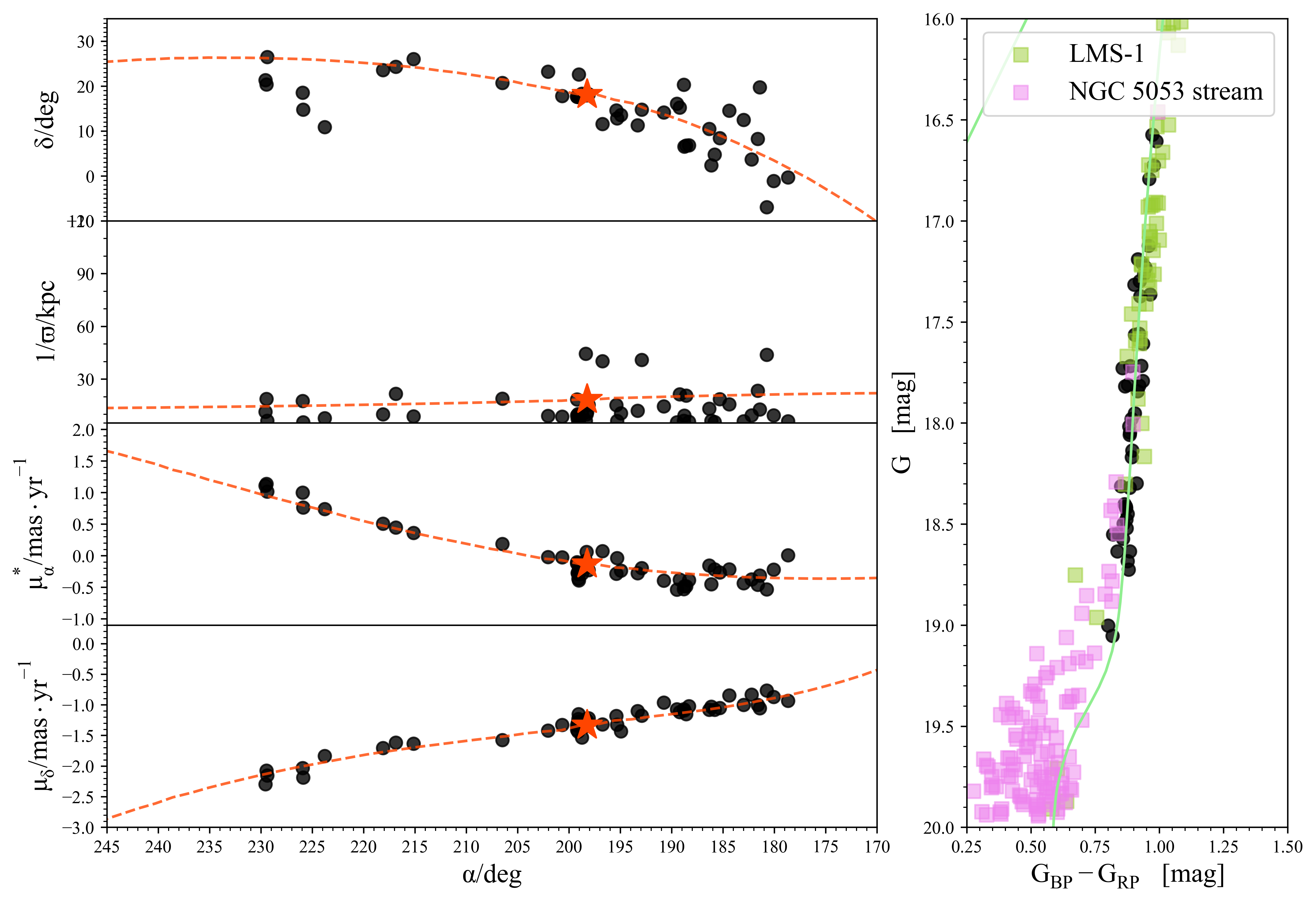}
  \caption{Proposed member candidates selected for the NGC~5024 stream. Left panel: Stream member candidates (black dots) shown in various phase-space projections, analogous to the left panel of Figure~\ref{fig:cmd_mock}. The red marker and dashed line indicate the central position of NGC~5024 and the median track of its mock stream, respectively. Right panel: Member stars plotted in the CMD with PARSEC isochrone overlaid as a green line. Known LMS-1 member stars \citep{2021ApJ...920...51M} and NGC~5053 stream members \citep{2026ApJS..283...60C} are also shown as green and purple squares, respectively.}
  \label{fig:members}
\end{figure*}

The simulated stream of NGC~5024 closely resembles those of the NGC~5053 and LMS-1 streams in $\alpha$, $\delta$, $\mu_{\alpha}^{*}$, $\mu_{\delta}$. Moreover, NGC~5024 and NGC~5053 share similar ages \citep{2021AJ....162...42W} and metallicities \citep[$\mathrm{[Fe/H]=-2.0}$ and $\mathrm{[Fe/H]=-2.17}$, respectively;][]{2020ApJ...900..146C}, while LMS-1 has a metallicity comparable with those of the two clusters \citep[$\mathrm{[Fe/H]=-2.1}$;][]{2021ApJ...920...51M}. Consequently, the NGC~5024 streams cannot be separated from NGC~5053 stream or LMS-1 in any of the phase-spaces projections displayed in Figure~\ref{fig:members}, including in a CMD.

\begin{figure*}[htbp!]
  \centering
  \includegraphics[width = 1.00\textwidth]{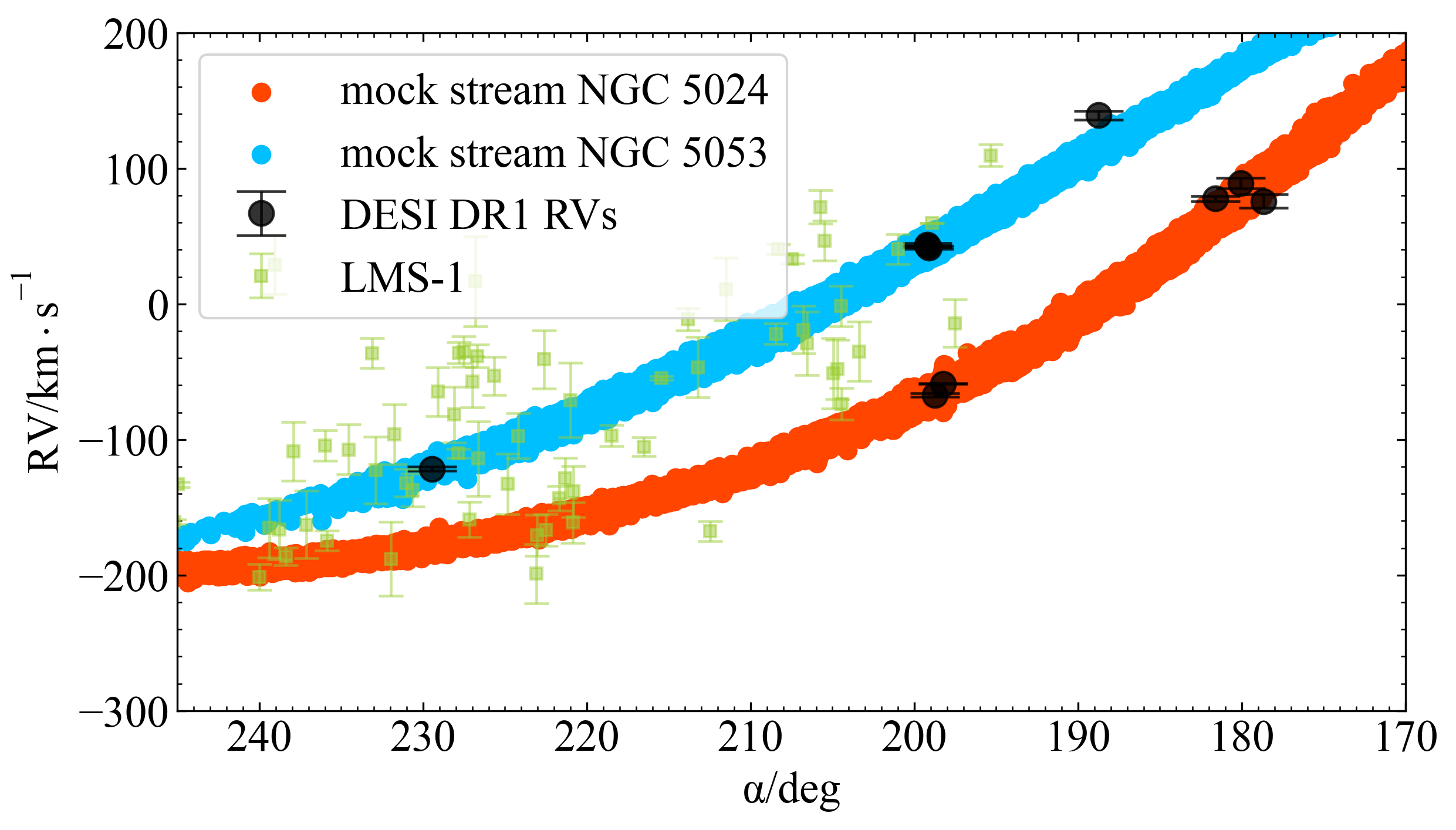}
  \caption{Our member candidates compared with simulated member stars of NGC~5024 (red) and NGC~5053 (blue) in the RV-$\alpha$ diagram. The black markers show the RVs and their uncertainties from DESI DR1 catalog. Stream stars belonging to LMS-1 are additionally plotted in green squares.}
  \label{fig:rvs}
\end{figure*}

However, we notice that the simulated RV track of NGC~5024 as a function of $\alpha$ is clearly separated from those of NGC~5053 and LMS-1, while NGC~5053 still follows a track very close to that of LMS-1 \citep{2021ApJ...920...51M}. Therefore, RV-$\alpha$ track can be effectively used to distinguish the tidal stream of NGC~5024 from those of NGC~5053 and LMS-1. Cross-matching our sample with DESI\footnote{Dark Energy Spectroscopic Instrument} DR1 \citep{2026AJ....171..285D,2026OJAp....955260K} yields $10$ common stars. We present their radial velocities in the RV-$\alpha$ plane, as shown in Figure~\ref{fig:rvs}. It is obvious that our member stars show two distinct RV-$\alpha$ tracks in the plot. All these stars with DESI DR1 RV measurements follow exactly either the RV-$\alpha$ track of the simulated stream of NGC~5024 or the track of NGC~5053 stream. The RV offset between NGC~5024 stream and the NGC~5053 (or LMS-1) stream reaches $\sim 90$ km s$^{-1}$ at $\alpha \approx 180^{\circ}$, far exceeding the internal RV dispersion of LMS-1 \citep[$\sigma_{\mathrm{RV}} = 20 \pm 4$ km s$^{-1}$;][]{2021ApJ...920...51M}. This large separation confirms that the two RV-$\alpha$ sequences provide robust evidence for the existence of a long stellar stream from NGC~5024. Based on DESI DR1 RVs, we confirm $5$ members of the NGC~5024 stream, which are labeled in the last column of Table~\ref{t:candidates}. The remaining $5$ stars are likely associated with LMS-1 or NGC~5053 streams, though distinguishing between these two streams remains difficult with the current data. In brief, the DESI RVs clearly illustrate that we have detected reliable member stars belonging to the trailing tail of NGC~5024, however, the sample is also contaminated by members of other streams, such as the NGC~5053 or LMS-1 streams.

\subsection{[Fe/H] from DESI}

\begin{figure*}[htbp!]
  \centering
  \includegraphics[width = 0.81\textwidth]{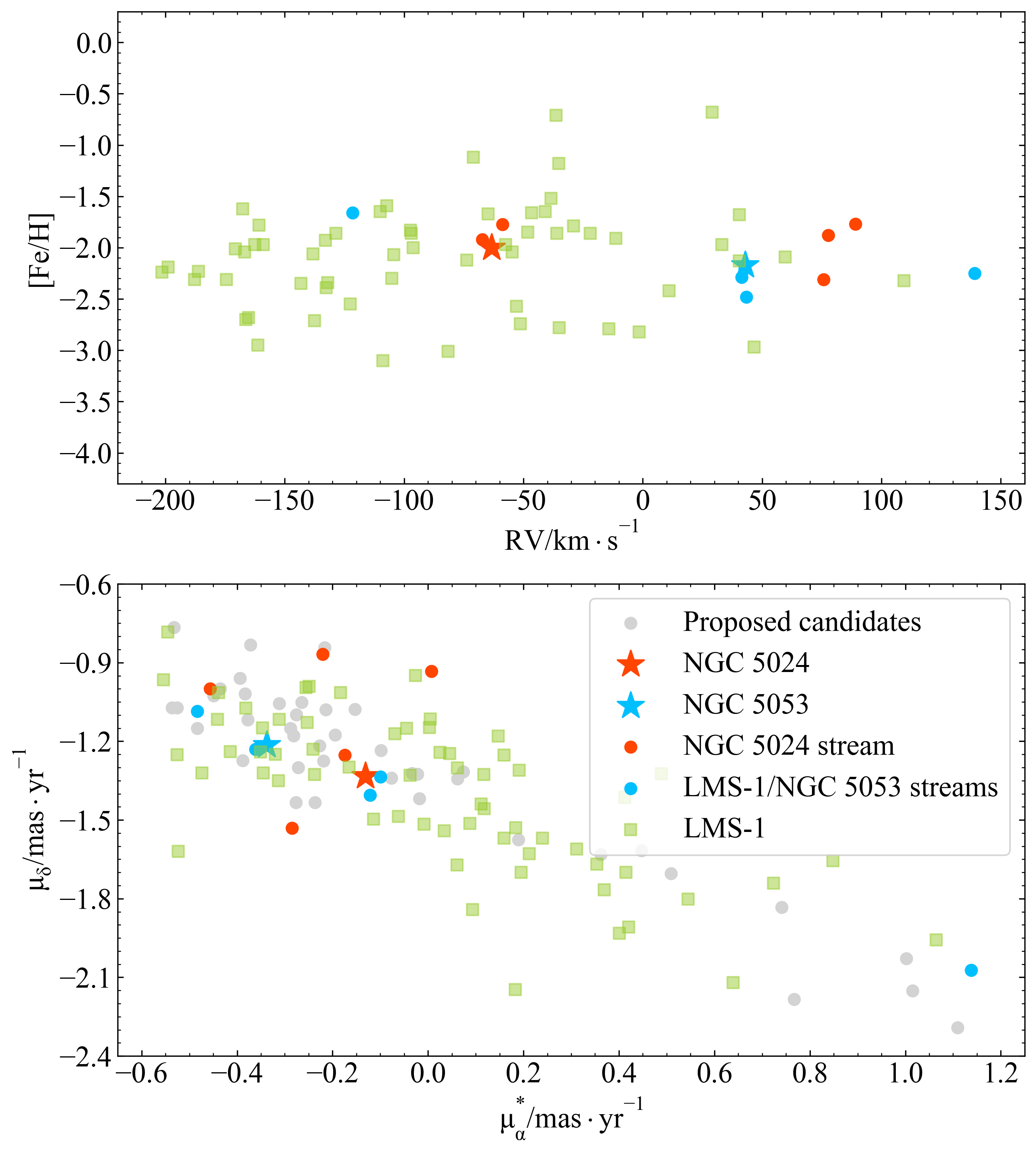}
  \caption{The [Fe/H]-RV plane and proper motion distributions for the GC NGC~5024 and NGC~5053, along with their stream members identified in this study, and for the LMS-1 stream members from the literature.}
  \label{fig:feh_rv_pms}
\end{figure*}

Furthermore, the mean metallicity of these candidates from DESI, as shown in Table~\ref{t:candidates} in Appendix~\ref{sec:appendix} ($\mathrm{[Fe/H]}=-2.05 \pm 0.27$), is also consistent with the metallicities of NGC~5024, NGC~5053, and LMS-1. In Figure~\ref{fig:feh_rv_pms}, we plot [Fe/H] against RV (from DESI DR1) for confirmed NGC~5024 stream members and for stars confirmed to be associated with the LMS-1/NGC~5053 streams, along with literature values for the two GCs and LMS-1 stream. Compared to LMS-1 (literature values), the NGC~5024 stream members (red dots in the top panel) show a significantly smaller [Fe/H] dispersion. The bottom panel presents the proper motions distributions in the $\mu_{\alpha}^{*}$-$\mu_{\delta}$ plane. Our proposed member candidates for the NGC~5024, NGC~5053, and LMS-1 streams fall in the same region as the LMS-1 stars reported in the literature. Therefore, the distinct RV-$\alpha$ sequence in the vicinity of these two GCs provides the most efficient-and likely the only-way to separate genuine NGC~5024 stream members from those belonging to LMS-1/NGC~5053.

\section{Summary}\label{sec:summary}

Based on Gaia DR3, we extract a data sample around NGC~5024. The sample is cleaned to ensure reliable astrometry and photometry, and only faint stars ($16.5 < G_{0} <20$) are considered in the search for tidal tails around this cluster. We use \texttt{Gala} to simulate a mock stream for NGC~5024 and employ the PARSEC isochrone to trace the cluster's locus in the CMD. A modified matched filter method is then adopted to identify its tidal stream. A long tidal stream extending around NGC~5024 from $\alpha \sim 230^{\circ}$ to $\alpha \sim 180^{\circ}$ is discovered. The leading tail overlaps with the LMS-1 stream member stars reported in the literature, while the trailing tail lies roughly $20^{\circ}$ beyond the end of LMS-1. We also explored different Galactic potentials (e.g., the potentials adopted in \citealt{2022MNRAS.513..853Y}) and isochrones with modest variations in metallicity and age. The long tidal feature persists, albeit with slight variations in the candidate star sample.

Because the detected stream is broad and the simulated streams of NGC~5024, NGC~5053, and LMS-1 show very similar properties in many phase-space projections (including [Fe/H]), we find that the RV track as a function of $\alpha$ is the most and probably only efficient way to distinguish genuine NGC~5024 member candidates from stars belonging to NGC~5053 and LMS-1. Out of our $51$ proposed member candidates, which are likely a mixture of stars belonging to NGC~5024, NGC~5053, and LMS-1, cross-matching with DESI DR1 yields $10$ common sources. These $10$ stars follow two distinct RV-$\alpha$ sequences: one ($5$ stars) matches the simulated stream of NGC~5024, and the other ($5$ stars) matches that of NGC~5053. We therefore report the discovery of extensive tidal tails, particularly the trailing tail, originating from NGC~5024, and we note that this stream lies partially embedded within the streams of NGC~5053 and LMS-1. Future spectroscopic follow-up of these member candidates, particularly in RV, will be crucial to disentangle the NGC~5024 stream from the overlapping NGC~5053 (LMS-1) tidal structures.

\begin{acknowledgments}

     This study is supported by the National Key R\&D Program of China under grant Nos. 2023YFE0107800 and 2024YFA1611900, and the National Natural Science Foundation of China under grant Nos. 12503024, 12273055, and 12588202. This study is also supported by International Partnership Program of Chinese Academy of Sciences grant No. 178GJZ2022040GC.
  
     XY acknowledges the support from the China Scholarship Council.

     JZ acknowledges financial support from the National Astronomical Observatories, CAS, under grant No. E4ZB0301.

     This work presents results from the European Space Agency (ESA) space mission Gaia. Gaia data are being processed by the Gaia Data Processing and Analysis Consortium (DPAC). Funding for the DPAC is provided by national institutions, in particular the institutions participating in the Gaia MultiLateral Agreement (MLA). The Gaia mission website is \url{https://www.cosmos.esa.int/gaia}. The Gaia archive website is \url{https://archives.esac.esa.int/gaia}.

     This research used data obtained with the Dark Energy Spectroscopic Instrument (DESI). DESI construction and operations is managed by the Lawrence Berkeley National Laboratory. This material is based upon work supported by the U.S. Department of Energy, Office of Science, Office of High-Energy Physics, under Contract No. DE–AC02–05CH11231, and by the National Energy Research Scientific Computing Center, a DOE Office of Science User Facility under the same contract. Additional support for DESI was provided by the U.S. National Science Foundation (NSF), Division of Astronomical Sciences under Contract No. AST-0950945 to the NSF’s National Optical-Infrared Astronomy Research Laboratory; the Science and Technology Facilities Council of the United Kingdom; the Gordon and Betty Moore Foundation; the Heising-Simons Foundation; the French Alternative Energies and Atomic Energy Commission (CEA); the National Council of Humanities, Science and Technology of Mexico (CONAHCYT); the Ministry of Science and Innovation of Spain (MICINN), and by the DESI Member Institutions: www.desi.lbl.gov/collaborating-institutions. The DESI collaboration is honored to be permitted to conduct scientific research on I’oligam Du’ag (Kitt Peak), a mountain with particular significance to the Tohono O’odham Nation. Any opinions, findings, and conclusions or recommendations expressed in this material are those of the author(s) and do not necessarily reflect the views of the U.S. National Science Foundation, the U.S. Department of Energy, or any of the listed funding agencies.

\end{acknowledgments}

\begin{contribution}

     XY led the data analysis, simulations, and calculations, and was responsible for writing and submitting the manuscript.

     YY provided helpful instructions and suggestions on simulations and calculations.

     JZ and HT provided helpful suggestions for the data analysis and manuscript editing.

     YC and GZ secured the financial support for the project leading to this publication,  and contributed to manuscript editing.

\end{contribution}


\software{
          \texttt{Agama} \citep{2019MNRAS.482.1525V},
          \texttt{Astropy} \citep{2013A&A...558A..33A,2018AJ....156..123A,2022ApJ...935..167A},
          \texttt{dustmaps} \citep{2018JOSS....3..695G},
          \texttt{Gala} \citep{gala,2024zndo..10449846P},
          \texttt{Matplotlib} \citep{2007CSE.....9...90H},
          \texttt{Numpy} \citep{2011CSE....13b..22V,2020Natur.585..357H},
          \texttt{Topcat} \citep{2005ASPC..347...29T}.
}

\appendix

\section{Candidate member stars of the NGC~5024, NGC~5053, and LMS-1 streams.}\label{sec:appendix}

Stars with the highest matched-filter weights for membership in the tidal tails of NGC~5024 are presented in Table~\ref{t:candidates}. We note, however, that the sample is also contaminated by stars from the LMS-1 and NGC~5053 streams. Follow-up spectroscopy to measure their RVs is essential for identifying which stream (NGC~5024, NGC~5053, or LMS-1) each star truly belongs to.

{
\setlength{\tabcolsep}{2.3pt}
\begin{longtable}{ccccccccc}
\caption{Proposed member candidates of the tidal tails of NGC~5024. Note: this sample is contaminated by stream stars belonging to NGC~5053 and LMS-1. The \texttt{source\underline{ }id}, right ascension (R.A.), declination (Decl.), proper motions $\left( \mu_{\alpha}^{*}, \mu_{\delta} \right)$ are taken from Gaia DR3, where the coordinates are at epoch J2016.0. The radial velocity (RV), the corresponding error ($\sigma_{\mathrm{RV}}$), and [Fe/H] are adopted from DESI DR1. In the last column, we list the likely progenitor for each star based on its RV measurement. \label{t:candidates}} \\
\hline
\texttt{source\underline{ }id} & R.A. (J2016) & Decl. (J2016) & $\mu_{\alpha}^{*}$ & $\mu_{\delta}$ & RV & $\sigma_{\mathrm{RV}}$ & [Fe/H] & Progenitor \\
    & deg & deg & mas $\cdot$ yr$^{-1}$ & mas $\cdot$ yr$^{-1}$ & km $\cdot$ s$^{-1}$ & km $\cdot$ s$^{-1}$ & dex & \\
\hline
\endfirsthead
\caption[]{Continued} \\
\hline
\texttt{source\underline{ }id} & R.A. (J2016) & Decl. (J2016) & $\mu_{\alpha}^{*}$ & $\mu_{\delta}$ & RV & $\sigma_{\mathrm{RV}}$ & [Fe/H] & Progenitor \\
    & deg & deg & mas $\cdot$ yr$^{-1}$ & mas $\cdot$ yr$^{-1}$ & km $\cdot$ s$^{-1}$ & km $\cdot$ s$^{-1}$ & dex & \\
\hline
\endhead
\hline \multicolumn{9}{r}{\small\emph{Continued on next page}} \\
\endfoot
\hline \hline
\endlastfoot
3794964868601766912 & 178.67 & -0.32 & 0.01 & -0.93 & 75.79 & 5.00 & -2.31 & NGC~5024 \\
3602739724618777472 & 180.06 & -1.12 & -0.22 & -0.87 & 89.08 & 3.77 & -1.77 & NGC~5024 \\
3594836774340055680 & 180.75 & -6.91 & -0.53 & -0.77 &   &   &   & \\
3951213644229912064 & 181.41 & 19.76 & -0.31 & -1.06 &   &   &   & \\
3904990969968309632 & 181.62 & 8.30 & -0.46 & -1.00 & 77.69 & 1.85 & -1.88 & NGC~5024 \\
3893751551655080576 & 182.20 & 3.72 & -0.37 & -0.83 &   &   &   & \\
3920436698860863872 & 182.99 & 12.54 & -0.44 & -1.00 &   &   &   & \\
3921183473415026432 & 184.39 & 14.58 & -0.22 & -0.84 &   &   &   & \\
3902150175519518720 & 185.32 & 8.48 & -0.26 & -1.05 &   &   &   & \\
3707961471272191488 & 185.81 & 4.78 & -0.21 & -1.08 &   &   &   & \\
3700637559896453248 & 186.13 & 2.41 & -0.45 & -1.03 &   &   &   & \\
3907022077182068864 & 186.33 & 10.48 & -0.15 & -1.08 &   &   &   & \\
3708700205647240064 & 188.32 & 6.88 & -0.38 & -1.02 &   &   &   & \\
3708689043028068224 & 188.58 & 6.78 & -0.48 & -1.15 &   &   &   & \\
3708617334254113152 & 188.76 & 6.54 & -0.48 & -1.09 & 139.07 & 3.31 & -2.25 & NGC~5053/LMS-1 \\
3949162986325093376 & 188.80 & 20.35 & -0.53 & -1.07 &   &   &   & \\
3933475566736899328 & 189.20 & 15.25 & -0.38 & -1.12 &   &   &   & \\
3935092467304792832 & 189.48 & 16.13 & -0.54 & -1.07 &   &   &   & \\
3932358703440734848 & 190.77 & 14.19 & -0.39 & -0.96 &   &   &   & \\
3930951642091273472 & 192.92 & 14.82 & -0.19 & -1.18 &   &   &   & \\
3735591824816997376 & 193.31 & 11.28 & -0.28 & -1.10 &   &   &   & \\
3929652466023499136 & 194.94 & 13.58 & -0.24 & -1.43 &   &   &   & \\
3737407359032957056 & 195.31 & 12.83 & -0.03 & -1.32 &   &   &   & \\
3930380651958545792 & 195.38 & 14.60 & -0.28 & -1.18 &   &   &   & \\
3736281798427250304 & 196.74 & 11.62 & 0.07 & -1.32 &   &   &   & \\
3938026243502900992 & 198.10 & 18.22 & -0.23 & -1.22 &   &   &   & \\
3939528760500517120 & 198.24 & 18.34 & -0.17 & -1.25 & -58.86 & 0.30 & -1.77 & NGC~5024 \\
3937264969139301760 & 198.30 & 18.00 & 0.06 & -1.34 &   &   &   & \\
3937267095145868672 & 198.36 & 18.07 & -0.08 & -1.34 &   &   &   & \\
3938771708089696640 & 198.37 & 18.13 & -0.10 & -1.24 &   &   &   & \\
3938778446893683328 & 198.74 & 18.32 & -0.29 & -1.53 & -67.25 & 1.14 & -1.92 & NGC~5024 \\
3938682750728482816 & 199.03 & 17.73 & -0.39 & -1.27 &   &   &   & \\
1445748182960326656 & 199.04 & 22.63 & -0.22 & -1.27 &   &   &   & \\
3938492294697450368 & 199.07 & 17.63 & -0.29 & -1.15 &   &   &   & \\
3938492088539015552 & 199.09 & 17.63 & -0.36 & -1.23 & 41.50 & 1.07 & -2.29 & NGC~5053/LMS-1 \\
3938492122898756480 & 199.10 & 17.63 & -0.28 & -1.43 &   &   &   & \\
3938494528080510720 & 199.17 & 17.72 & -0.27 & -1.30 &   &   &   & \\
3938688694961600512 & 199.20 & 17.79 & -0.10 & -1.33 & 43.41 & 1.61 & -2.48 & NGC~5053/LMS-1 \\
3938493398504321280 & 199.21 & 17.66 & -0.12 & -1.41 & 42.72 & 0.40 & -2.21 & NGC~5053/LMS-1 \\
3938549709819621760 & 200.65 & 17.84 & -0.02 & -1.33 &   &   &   & \\
1443177254192367744 & 202.05 & 23.26 & -0.02 & -1.42 &   &   &   & \\
1249375990355153536 & 206.48 & 20.75 & 0.19 & -1.58 &   &   &   & \\
1259320385833294720 & 215.14 & 26.04 & 0.36 & -1.63 &   &   &   & \\
1254792188928956544 & 216.85 & 24.31 & 0.45 & -1.62 &   &   &   & \\
1254825586594609408 & 218.12 & 23.57 & 0.51 & -1.70 &   &   &   & \\
1180320063060840064 & 223.79 & 10.90 & 0.74 & -1.83 &   &   &   & \\
1183723223347647488 & 225.88 & 14.80 & 0.77 & -2.18 &   &   &   & \\
1188646767697698816 & 225.95 & 18.58 & 1.00 & -2.03 &   &   &   & \\
1271194149340913280 & 229.38 & 26.49 & 1.02 & -2.15 &   &   &   & \\
1212570285292430848 & 229.46 & 20.38 & 1.14 & -2.07 & -121.63 & 1.48 & -1.66 & NGC~5053/LMS-1 \\
1214889743726083200 & 229.56 & 21.35 & 1.11 & -2.29 &   &   &   & \\
\end{longtable}
}

\bibliography{bibliography}{}
\bibliographystyle{aasjournalv7}

\end{document}